\documentclass[aps,prl,twocolumn,superscriptaddress,showpacs]{revtex4-1}
\usepackage{graphicx}
\def\RM#1{}

\begin{document}

\title{Stick-Slip Motion of the Wigner Solid on Liquid Helium}


\author{David G. Rees}
\email[]{drees@nctu.edu.tw}
\affiliation{NCTU-RIKEN Joint Research Laboratory, Institute of Physics, National Chiao Tung University, Hsinchu 300, Taiwan}
\affiliation{RIKEN CEMS, Wako 351-0198, Japan}

\author{Niyaz R. Beysengulov}
\affiliation{RIKEN CEMS, Wako 351-0198, Japan}
\affiliation{KFU-RIKEN Joint Research Laboratory, Institute of Physics, Kazan Federal University, Kazan, Russia}

\author{Juhn-Jong Lin}
\affiliation{NCTU-RIKEN Joint Research Laboratory, Institute of Physics, National Chiao Tung University, Hsinchu 300, Taiwan}
\affiliation{RIKEN CEMS, Wako 351-0198, Japan}
\affiliation{Department of Electrophysics, National Chiao Tung University, Hsinchu 300, Taiwan}

\author{Kimitoshi Kono}
\affiliation{NCTU-RIKEN Joint Research Laboratory, Institute of Physics, National Chiao Tung University, Hsinchu 300, Taiwan}
\affiliation{RIKEN CEMS, Wako 351-0198, Japan}
\affiliation{KFU-RIKEN Joint Research Laboratory, Institute of Physics, Kazan Federal University, Kazan, Russia}



\date{\today}

\begin{abstract}
  We present time-resolved transport measurements of a Wigner solid
  (WS) on the surface of liquid Helium confined in a micron-scale
  channel. At rest, the WS is `dressed' by a cloud of quantised
  capillary waves (ripplons). Under a driving force, we find that
  repeated WS-ripplon decoupling leads to stick-slip current
  oscillations, the frequency of which can be tuned by adjusting the
  temperature, pressing electric field, or electron density. The WS on
  liquid He is a promising system for the study of polaron-like
  decoupling dynamics.
\end{abstract}

\maketitle

Surface-state electrons (SSEs) on liquid helium substrates (Fig. 1)
form a model Coulomb system\cite{Andrei}, the ground state of which
remains the clearest example of the classical Wigner solid
(WS)\cite{GrimesAdamsWignerCrystal}. The WS is dressed by a cloud of
ripplons, the Bragg scattering of which from the electron lattice
gives rise to a commensurate deformation of the He surface known as
the dimple lattice
(DL)\cite{MonarkhaShikinDimple,Fisher1979,saitoh1986,leiderer1982macroscopic}. The WS-DL
system is analogous to polaron states in which electrons are dressed
by a cloud of virtual phonons, or lattice
deformation\cite{alexandrov2010advances,dykman2015roots,PhysRevLett.52.1449}. On
liquid He individual electrons do not perform self-trapping; rather,
the DL appears as a consequence of the long-range electron
ordering. Also, the lattice constant of the WS (and so the DL) is no less than 100 nm and a continuum description of the medium (liquid He)
is applicable. Despite these differences, the decoupling dynamics of
the WS and polaron systems exhibit strong
similarities\cite{PhysRevB.69.235205,ge1998femtosecond},
although the rate depends on the different binding energies; polaron
decoupling in a GaAs crystal occurs in the femtosecond
range\cite{gaal2007}, whereas the decoupling of the WS from the DL
takes place on nanosecond timescales. It is therefore more
straightforward to observe the real-time decoupling for the WS than
for polarons.

In contrast with other systems, the coupling of the WS with the DL is of a dynamical nature, due to a resonant interaction between the WS and ripplons emitted by the moving electron lattice. According to hydrodynamic theory, when the WS is at rest the depth of the surface dimples $\xi_0$ can be estimated as 
\begin{eqnarray}  \label{eq:dimple}
\xi_0\approx -\frac{\sqrt{3}eE_z}{8\pi^2\sigma}e^{-W}~~,
\end{eqnarray}

where $e$, $E_z$, $\sigma$ and $e^{-W}$ are the elementary charge, perpendicular pressing electric field, surface tension coefficient of liquid He and the self-consistent Debye-Waller factor, respectively\cite{MKDimple}. Typically, $\xi_0\approx10^{-12}$ m. However, when the velocity of the WS-DL system approaches the phase velocity of ripplons of wavevector equal to the WS periodicity, $v_{BC}$, constructive interference resonantly deepens the DL. As a result, the resistive force exerted on the WS increases dramatically. This is called the Bragg-Cherenkov (BC) effect\cite{DykmanRubo,Vinen,Kristensen1996}. The decoupling of the WS therefore occurs from this dynamically pinned state and, hence, under strongly nonequilibrium conditions. So far, however, very little is known about this decoupling process. It is an interesting problem concerning strongly correlated systems far from equilibrium, and also from a hydrodynamics point of view.

\begin{figure}[b]
\begin{centering}
 \includegraphics[angle=0,width=0.45\textwidth]{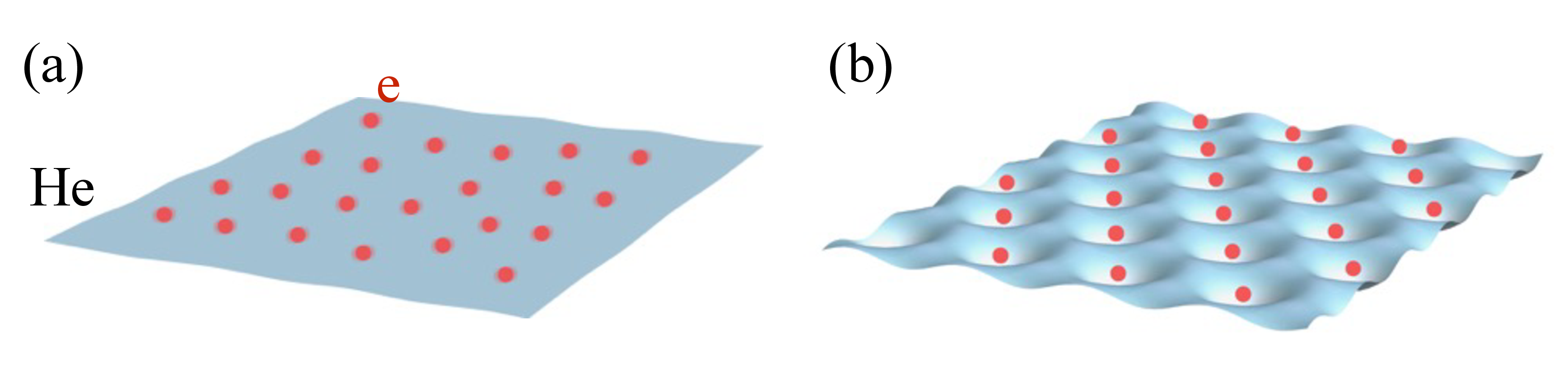}
\caption{(Color online) Schematic depiction of electrons on the surface of liquid helium. (a) Above $T_m$ electrons move freely in the 2D plane above the surface. (b) Below $T_m$ the Wigner solid is dressed by capillary waves, giving rise to the dimple lattice.\label{Fig:1}}
\end{centering}
\end{figure}

The WS decoupling from the DL was first observed in Corbino
  conductivity measurements under perpendicular magnetic
  fields\cite{PhysRevLett.74.781}.  The decoupling was signalled by an abrupt jump in conductivity on sweeping the amplitude or frequency of the
  sinusoidal driving potential, or the magnetic field, or so
  on\cite{Shirahama2}. In such experiments, magnetic fields are necessary because an electric field large enough to cause the decoupling
  cannot be applied to the WS due to its high mobility.  The electric current then flows azimuthally, whereas the
  electric field acts in the radial direction.  The Hall angle is
  almost 90 degrees, and hence, the azimuthal current or velocity of the SSEs is extremely difficult to determine.  Furthermore, the
  electric field is spatially inhomogeneous and time varying.  These
  ambiguities make analysis of the decoupling process difficult.

Recently, SSEs in capillary-condensed microchannel
  devices have been found to show the decoupling phenomena without applying
  magnetic fields\cite{Ikegami2009}. In such devices the electric current is homogeneous but is typically measured using a sinusoidal driving voltage. The transport measurement is then inevitably complicated by the nonlinear response of the WS during each ac cycle. In this Letter we report simultaneous measurements of the WS
  velocity and driving electric field by employing a
  linear sweep of the driving electric potential.  That is, we performed
  the first time-resolved transport measurement of a quasi-1D WS confined in a
  microchannel.  We demonstrate that, under a driving potential,
  repeated ripplon dressing and decoupling causes stick-slip motion of
  the WS and so oscillations of the electron velocity. Our experiment allows the control
  and quantitative analysis of the decoupling dynamics of the WS.

\begin{figure} 
\begin{centering}
 \includegraphics[angle=0,width=0.35\textwidth]{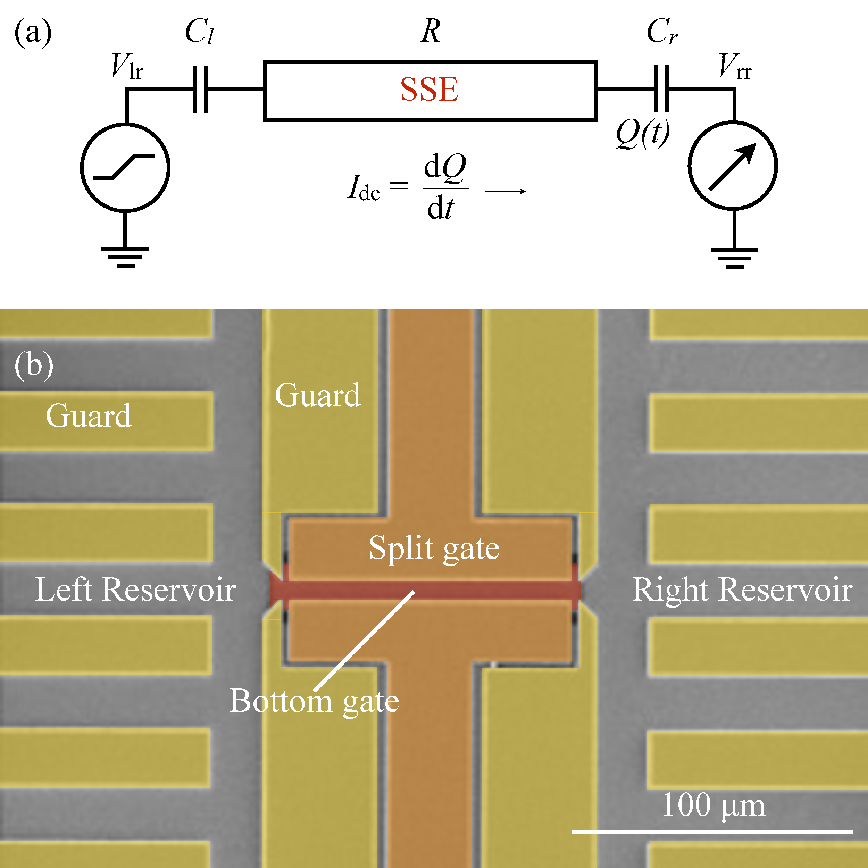}
\caption{(Color online) (a) Schematic circuit diagram of the dc current measurement. The SSE in the CM form an electrical resistance $R$. The electron system is coupled to the external circuit by the capacitances $C_l$ and $C_r$. The current $I_{dc}$ flows through the CM in response to varying $V_{lr}$, resulting in the transfer of charge $Q(t)$ from the left to the right reservoir, which accumulates in $C_l$ and $C_r$. The capacitive coupling to the Guard electrode is omitted for simplicity. (b) False-colour scanning electron micrograph of a sample identical to those used in the experiment. The Left and Right Reservoirs each contain 25 microchannels of length 0.7 mm connected in parallel. \label{Fig:1}}
\end{centering}
\end{figure}

The device used in the experiment is shown in Fig. 2(b). Our sample is electrons floating on a helium surface in a microchannel 7.5 $\mu$m wide, 100 $\mu$m long and $h=2.2$ $\mu$m deep. We denote this the central microchannel (CM). Two large arrays of microchannels, which act as electron reservoirs, are attached to the CM. The SSE in the reservoirs are capacitively coupled to the external circuit by $C_l$ and $C_r$, as shown schematically in Fig. 2(a). To measure SSE transport through the CM, the voltage $V_{lr}$ applied to the electrode beneath the He in the left reservoir (Left Reservoir electrode) was ramped from 0 to +50 mV in a time $t_{r}$. As $V_{lr}$ is varied, charge moves between the reservoirs. After the ramp period, the charge distribution achieves a new equilibrium state and the current stops. To determine the charge accumulated in $C_l$ and $C_r$, and so the SSE current $I_{dc}$ passing through the CM, the current flowing in the electrode beneath the He in the right reservoir (Right Reservoir electrode, $V_{rr}=0$ V) was recorded. To tune the electrostatic confinement for SSEs in the CM, voltages $V_{bg}$ and $V_{sg}$ were applied to the electrode beneath the He in the CM (Bottom Gate electrode) and the electrode in the plane of the He surface (Split Gate electrode), respectively. The voltage on the Guard electrode was -0.2 V. 

The current measurement was performed using a current preamplifier and a digital storage oscilloscope. The measurement was averaged over several thousand cycles. A small current component due to 
cross-talk between the electrical cables was subtracted from the measurement. The bandwidth of the preamplifier was 200 kHz, which leads to the smoothing of some of the transport features shown here but does not detract from the essential physics. Finite-element modelling (FEM)\cite{hecht2012new} was used to calculate the average areal electron density $n_s$ (m$^{-2}$), the electrostatic potential of the electron system $V_e$ (which depends on $n_s$), the effective width $w_{e}$ of the charge sheet in the CM, the linear electron density $n_{l}$ (m$^{-1}$), and the number of electron rows $N_{y}$, for varying bias conditions\cite{beysengulov2016structural,Supp}.  

\begin{figure} 
\begin{centering}
  \includegraphics[angle=0,width=0.35\textwidth]{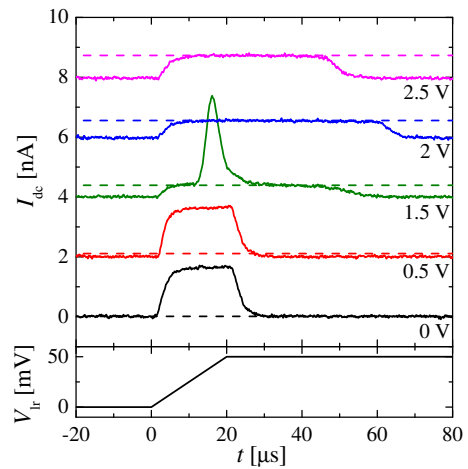}
 \caption{(Color online) $I_{dc}$ against time $t$ for $V_{sg}=-0.2$ V and different values of $V_{bg}$ as indicated on the plot. Here $T=0.6$ K. Each data set is shifted vertically by 2 nA for clarity. The dashed lines indicate the value of $I_{BC}$ expected for each value of $V_{bg}$. The lower panel shows the $V_{lr}$ voltage waveform. \label{Fig:2}}
\end{centering}
\end{figure}

Results of transient current measurements are shown in Fig. 3 for $t_{r}=20$ $\mu$s, $V_{sg}=-0.2$ V. Generally, current flows in the CM in response to the $V_{lr}$ ramp. However, the transport is strongly influenced by $V_{bg}$, the value of which determines the electron density in the CM. The WS forms when the ratio of the electron Coulomb energy to kinetic energy $\Gamma=E_C/E_k=e^2\sqrt{\pi n_s}/4\pi\varepsilon_0 k_B T$ reaches a threshold value of $\sim130$, where $\varepsilon_0$, $k_B$ and $T$ are the vacuum permittivity, the Boltzmann constant and the temperature, respectively\cite{GrimesAdamsWignerCrystal}. At the experimental temperature $T=0.6$ K the WS should form when $n_s\approx7.1\times10^{12}$ m$^{-2}$, which occurs when $V_{bg}=0.18$ V according to the FEM\cite{Supp}. It is clear that for both the electron liquid ($V_{bg}=0$ V) and the low-density WS ($V_{bg}=0.5$ V) a constant current flows in the CM as $V_{lr}$ changes. For $V_{bg}=1.5$ V the current is initially suppressed but then suddenly increases, during the $V_{lr}$ ramp, before relaxing. For $V_{bg}=2.0$ V and 2.5 V, $I_{dc}$ is small and continues to flow after the ramp phase.

Because the SSE system is capacitively coupled to both the Reservoir and Guard electrodes, the measurement of the current registered on the Right Reservoir electrode does not give the true current flowing in the CM. The average charge detected by integrating the current measurements shown in Fig. 3 is $Q^{av}=1.58\times10^5\cdot e$ with standard deviation $0.04\times10^5\cdot e$. The FEM analysis gives the expected value $Q^{FEM}=2.03\times10^5\cdot e$. The deficit is due to the displacement current flowing into the Guard electrode. In Figs. 3 and 4, the current measurement is therefore corrected by the factor $Q^{FEM}/Q^{av} =1.28$. 

BC scattering followed by WS-DL decoupling results in an abrupt jump in the WS conductivity with increasing driving force. The influence of BC scattering on the WS velocity in the CM can be estimated using the periodicity of the WS along the channel, $a=N_{y}/n_{l}$. The limiting velocity is given by $v_{BC}=(\sigma k/\rho)^{1/2}$, where the wavevector $k=2\pi/a$ and $\rho$ is density of liquid He. The typical electron spacing is 300 nm, and therefore $kh\gg1$. In this `thick-film' limit the dispersion relation for ripplons on bulk liquid He $\omega^2=(\sigma/\rho)k^3$ is applicable. The BC scattering-limited current was calculated as $I_{BC}=n_lev_{BC}$ for the bias conditions in Figs. 3 and 4. 

On comparing $I_{BC}$ with the measurements in Fig. 3, it is clear that when $V_{bg}$ is sufficiently large the WS is pinned to the DL and, due to BC scattering, $I_{dc}=I_{BC}$. Then, because $I_{dc}$ is limited, the potential difference between the left and right reservoirs builds up during the $V_{lr}$ ramp and the current continues to flow after the ramp phase. For $V_{bg}=1.5$ V the electron velocity is initially limited by the BC scattering but, as the potential difference builds, suddenly increases when the electrons decouple from the DL. This releases the force and the electron velocity quickly returns to $v_{BC}$. For $V_{bg}=0.5$ V the driving force always overcomes the weaker pinning and the current exceeds $I_{BC}$.

The stronger pinning of the WS with increasing $V_{bg}$ is expected, as $n_{s}$ (and so $\Gamma$) and $E_{z}$ both increase with $V_{bg}$. However, by keeping the electrode bias conditions fixed and changing temperature, the influence of the electron thermal motion on the WS decoupling can be determined unambiguously. Approaching the WS melting point, thermal fluctuations in the electron positions should degrade the DL, leading to a decrease in the decoupling threshold force. This in turn should lead to a decrease in the time taken for the decoupling to occur in our transport measurements. In Fig. 4(a) we show $I_{dc}$ against time $t$ at different temperatures. Here $t_{r}= 80$ $\mu$s. The Split-Gate voltage was $V_{sg}=-1$ V which reduces $n_{s}$ and $w_{e}$. This increases the resistance of the SSE system $R$ and thereby slows the system response to the $V_{lr}$ ramp (see Fig. 2(a)). During the ramp phase, multiple decoupling events occur, at regular intervals. After each decoupling event $I_{dc}$ returns to $I_{BC}$ and the stick-slip cycle is repeated. For increasing $T$ (decreasing $\Gamma$) the stick-slip period decreases as the time taken to reach the decoupling threshold force is reduced. 

Similar narrow-band current oscillations, also attributed to WS sliding, have been observed for degenerate electron systems in the WS regime\cite{Csathy}. However, for such cases little is known about the nature of the collective electron ground states or the mechanisms by which pinning forces are overcome. Qualitatively similar phenomena are observed in charge/spin density wave systems. But, contrary to those systems, the present system does not have a pinning mechanism due to irregularities. In our experiments, the quantitative understanding of the BC scattering mechanism allows us to clearly demonstrate the link between WS sliding and the appearance of spontaneous current oscillations.

\begin{figure} 
\begin{centering}
\includegraphics[angle=0,width=0.48\textwidth]{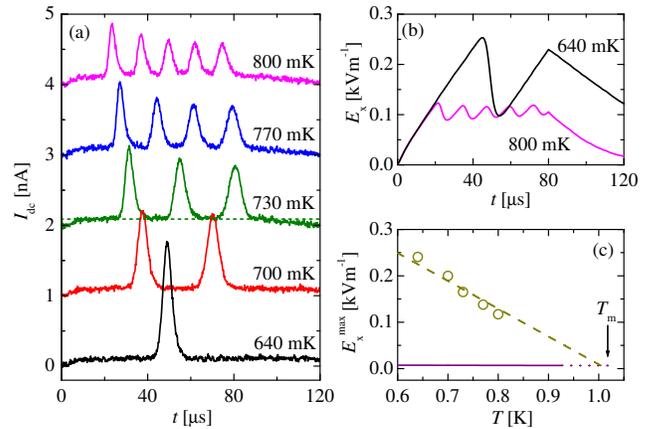}
\caption{(Color online) (a) $I_{dc}$ against time $t$ at different temperatures from 640 mK to 800 mK, as indicated on the plot. Here $t_{r}=80$ $\mu$s, $V_{bg}=1.2$ V and $V_{sg}=-1$ V. Each data set is shifted by 1 nA for clarity. For $T=730$ mK, we plot the expected $I_{BC}$ (dashed line). (b) $E_x$ against $t$ for the data shown in (a) for $T=640$ mK and 800 mK. (c) $E_x^{max}$ against $T$, for the data shown in (a). The dashed line is a guide to the eye. The solid/dotted line indicates the value of $E_x^{max}$ expected for a dimple depth $\xi_0$. $T_{m}=1.02$ K is indicated by the arrow.\label{Fig:3}}
\end{centering}
\end{figure}

\begin{figure} 
\begin{centering}
\includegraphics[angle=0,width=0.45\textwidth]{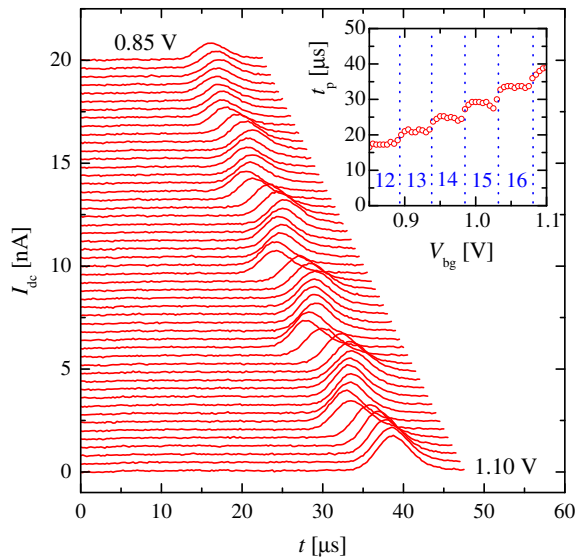}
\caption{(Color online) $I_{dc}$ against $t$ for different values of $V_{bg}$. $V_{bg}$ is varied from 1.10 to 0.85 V in 5 mV steps. Each data set is shifted by 0.4 nA for clarity. Inset: the corresponding values of the $I_{dc}$ peak position, $t_p$, against $V_{bg}$. The dotted lines mark the values of $V_{bg}$ at which structural transitions between electron row configurations occur, as calculated by the FEM. The values of $N_y$ between each transition are indicated.\label{Fig:3}}
\end{centering}
\end{figure}

Stick-slip motion typically results in a saw-tooth-type force-velocity profile\cite{persson2000sliding}. For the data shown in Fig. 4(a), to calculate the time-varying electric field along the CM, $E_x(t)$, we integrate $I_{dc}$ at each value of $t$ to obtain the charge accumulated in the right reservoir $Q(t)$. Analysis of the circuit diagram in Fig. 2(a) gives \mbox{$E_{x}(t)=[V_{lr}(t)-V_{rr}-2Q(t)/C]/l$}, where $l=100$ $\mu$m and $C=C_l=C_r=1.01$ pF is given by the measurement of $Q_{av}$. In Fig. 4(b) we show the dependence of $E_{x}$ on $t$, at $T=800$ and 640 mK. The driving field builds steadily when the WS is pinned by the DL and then drops rapidly as the WS decouples. The maximum value of the driving field $E_x^{max}$ is higher at lower temperature, as expected. For this measurement, the FEM gives $n_{s}=2.04\times10^{13}$ m$^{-2}$, for which the melting temperature $T_{m}=1.02$ K. As shown in Fig. 4(c), $E_x^{max}$ extrapolates close to this value, as the decoupling force becomes zero when the WS melts. 

Assuming a sinusoidal surface profile, equating the components of the forces acting on each electron in the direction of motion yields the expression $k\xi_{th}=E_x^{max}/E_z$, where $ \xi_{th}$ is the dimple depth at the decoupling threshold. The $E_x^{max}$ expected for the static DL ($\xi_{th}=\xi_0$) is shown in Fig. 4(c). (Note that the dependence of $\xi_0$ on $T$ is weak, without taking the WS melting into account). As expected, the experimental values of $E_x^{max}$ are larger than those for the static DL, as a consequence of the resonant deepening of the DL due to BC scattering. For $T=0.64$ K, we estimate $E_{x}^{max}/E_z=1.4\times10^{-3}$, $a=240$ nm, and so $\xi_{th}=5.2\times10^{-11}$ m. This value exceeds the static dimple depth $\xi_{0}\approx2\times10^{-12}$ m, and is consistent with other measurements\cite{Ikegami2009}.

For the bulk 2D WS the magnitude of the WS-DL coupling is estimated by $eE_z\xi_0$ (Eq. (1)).  Since $E_z$ and W are functions of  $V_{bg}$, the decoupling threshold depends on  $V_{bg}$, yet the dependence should be monotonous and smooth. In Fig. 5 we show $I_{dc}$ against $t$, recorded for $V_{bg}$ values between 0.85 and 1.10 V. Here $t_{r}=80$ $\mu$s, $V_{sg}=-1$ V and $T=0.6$ K. The time values at which $I_{dc}$ reaches its peak value, $t_p$, are plotted in the inset. Although the general trend, the increase of the decoupling threshold, is as expected, it is remarkably tortuous. We ascribe this effect to successive structural transitions of electron row formation. The numbers of electron rows along the quasi-1D channel, as calculated by the FEM, are indicated in the inset of Fig. 5. Clear correspondence is observed. Previous (ac) measurements have shown a similar modulation effect due to the reduced positional order of the WS at each structural boundary\cite{beysengulov2016structural, PhysRevLett.109.236802, *rees2013reentrant}. Further details of the influence of confinement on the WS ordering will be reported elsewhere.

Stick-slip motion is common in nature, although complex microscopic processes often determine macroscopic motion\cite{RevModPhys.85.529,lee2010frictional,lee2015noncontact}. Here, for SSEs in a microchannel confinement, we have demonstrated a quantitative understanding of stick-slip friction at a microscopic level and that the coupling between the WS and the He substrate depends on the positional ordering of the quasi-1D electron system. Recently, quasi-1D cold ion systems comprising less than 10 particles have been used to investigate the influences of dimensionality and commensurability on sliding friction\cite{Bylinskii}. Our results demonstrate that similar investigations can be conducted with quasi-1D SSE systems, which can contain a much larger number of particles, by measuring decoupling from the DL or from lithographically structured substrates. 

In conclusion, we have presented the first demonstration of stick-slip motion of the WS on a liquid He substrate. Our microchannel device allows precise control and quantitative analysis of the WS decoupling, which exhibits basic similarities with polaron dynamics. SSEs on He are a promising system to study stick-slip friction at the nanoscale.

\begin{acknowledgments}
  We are grateful to M. I. Dykman for helpful discussions.  This work was
  supported by JSPS KAKENHI Grant No. 24000007, and by the Taiwan
  Ministry of Science and Technology (MOST) through Grant Nos. NSC 102- 2112-M-009-015 and MOST 103-2112-M-009-017 and the the MOE ATU
  Program. This work was performed according to the Russian Government Program of Competitive Growth of Kazan Federal University.
\end{acknowledgments}

\bibliography{dctransport}

\begin{thebibliography}{30}%
\makeatletter
\providecommand \@ifxundefined [1]{%
 \@ifx{#1\undefined}
}%
\providecommand \@ifnum [1]{%
 \ifnum #1\expandafter \@firstoftwo
 \else \expandafter \@secondoftwo
 \fi
}%
\providecommand \@ifx [1]{%
 \ifx #1\expandafter \@firstoftwo
 \else \expandafter \@secondoftwo
 \fi
}%
\providecommand \natexlab [1]{#1}%
\providecommand \enquote  [1]{``#1''}%
\providecommand \bibnamefont  [1]{#1}%
\providecommand \bibfnamefont [1]{#1}%
\providecommand \citenamefont [1]{#1}%
\providecommand \href@noop [0]{\@secondoftwo}%
\providecommand \href [0]{\begingroup \@sanitize@url \@href}%
\providecommand \@href[1]{\@@startlink{#1}\@@href}%
\providecommand \@@href[1]{\endgroup#1\@@endlink}%
\providecommand \@sanitize@url [0]{\catcode `\\12\catcode `\$12\catcode
  `\&12\catcode `\#12\catcode `\^12\catcode `\_12\catcode `\%12\relax}%
\providecommand \@@startlink[1]{}%
\providecommand \@@endlink[0]{}%
\providecommand \url  [0]{\begingroup\@sanitize@url \@url }%
\providecommand \@url [1]{\endgroup\@href {#1}{\urlprefix }}%
\providecommand \urlprefix  [0]{URL }%
\providecommand \Eprint [0]{\href }%
\providecommand \doibase [0]{http://dx.doi.org/}%
\providecommand \selectlanguage [0]{\@gobble}%
\providecommand \bibinfo  [0]{\@secondoftwo}%
\providecommand \bibfield  [0]{\@secondoftwo}%
\providecommand \translation [1]{[#1]}%
\providecommand \BibitemOpen [0]{}%
\providecommand \bibitemStop [0]{}%
\providecommand \bibitemNoStop [0]{.\EOS\space}%
\providecommand \EOS [0]{\spacefactor3000\relax}%
\providecommand \BibitemShut  [1]{\csname bibitem#1\endcsname}%
\let\auto@bib@innerbib\@empty
\bibitem [{\citenamefont {Andrei}(1997)}]{Andrei}%
  \BibitemOpen
  \bibinfo {editor} {\bibfnamefont {E.~Y.}\ \bibnamefont {Andrei}},\ ed.,\
  \href@noop {} {\emph {\bibinfo {title} {Two-dimensional electron systems on
  helium and other cryogenic substrates.}}}\ (\bibinfo  {publisher} {Kluwer
  Academic, Dordrecht},\ \bibinfo {year} {1997})\BibitemShut {NoStop}%
\bibitem [{\citenamefont {Grimes}\ and\ \citenamefont
  {Adams}(1979)}]{GrimesAdamsWignerCrystal}%
  \BibitemOpen
  \bibfield  {author} {\bibinfo {author} {\bibfnamefont {C.~C.}\ \bibnamefont
  {Grimes}}\ and\ \bibinfo {author} {\bibfnamefont {G.}~\bibnamefont {Adams}},\
  }\href {\doibase 10.1103/PhysRevLett.42.795} {\bibfield  {journal} {\bibinfo
  {journal} {Phys. Rev. Lett.}\ }\textbf {\bibinfo {volume} {42}},\ \bibinfo
  {pages} {795} (\bibinfo {year} {1979})}\BibitemShut {NoStop}%
\bibitem [{\citenamefont {Monarkha}\ and\ \citenamefont
  {Shikin}(1975)}]{MonarkhaShikinDimple}%
  \BibitemOpen
  \bibfield  {author} {\bibinfo {author} {\bibfnamefont {Y.~P.}\ \bibnamefont
  {Monarkha}}\ and\ \bibinfo {author} {\bibfnamefont {V.~B.}\ \bibnamefont
  {Shikin}},\ }\href@noop {} {\bibfield  {journal} {\bibinfo  {journal} {Sov.
  Phys. JETP}\ }\textbf {\bibinfo {volume} {41}},\ \bibinfo {pages} {710}
  (\bibinfo {year} {1975})}\BibitemShut {NoStop}%
\bibitem [{\citenamefont {Fisher}\ \emph {et~al.}(1979)\citenamefont {Fisher},
  \citenamefont {Halperin},\ and\ \citenamefont {Platzman}}]{Fisher1979}%
  \BibitemOpen
  \bibfield  {author} {\bibinfo {author} {\bibfnamefont {D.~S.}\ \bibnamefont
  {Fisher}}, \bibinfo {author} {\bibfnamefont {B.~I.}\ \bibnamefont
  {Halperin}}, \ and\ \bibinfo {author} {\bibfnamefont {P.~M.}\ \bibnamefont
  {Platzman}},\ }\href {\doibase 10.1103/PhysRevLett.42.798} {\bibfield
  {journal} {\bibinfo  {journal} {Phys. Rev. Lett.}\ }\textbf {\bibinfo
  {volume} {42}},\ \bibinfo {pages} {798} (\bibinfo {year} {1979})}\BibitemShut
  {NoStop}%
\bibitem [{\citenamefont {Saitoh}(1986)}]{saitoh1986}%
  \BibitemOpen
  \bibfield  {author} {\bibinfo {author} {\bibfnamefont {M.}~\bibnamefont
  {Saitoh}},\ }\href@noop {} {\bibfield  {journal} {\bibinfo  {journal} {J.
  Phys. Soc. Jpn.}\ }\textbf {\bibinfo {volume} {55}},\ \bibinfo {pages} {1311}
  (\bibinfo {year} {1986})}\BibitemShut {NoStop}%
\bibitem [{\citenamefont {Leiderer}\ \emph {et~al.}(1982)\citenamefont
  {Leiderer}, \citenamefont {Ebner},\ and\ \citenamefont
  {Shikin}}]{leiderer1982macroscopic}%
  \BibitemOpen
  \bibfield  {author} {\bibinfo {author} {\bibfnamefont {P.}~\bibnamefont
  {Leiderer}}, \bibinfo {author} {\bibfnamefont {W.}~\bibnamefont {Ebner}}, \
  and\ \bibinfo {author} {\bibfnamefont {V.~B.}\ \bibnamefont {Shikin}},\
  }\href@noop {} {\bibfield  {journal} {\bibinfo  {journal} {Surface Science}\
  }\textbf {\bibinfo {volume} {113}},\ \bibinfo {pages} {405} (\bibinfo {year}
  {1982})}\BibitemShut {NoStop}%
\bibitem [{\citenamefont {Alexandrov}\ and\ \citenamefont
  {Devreese}(2010)}]{alexandrov2010advances}%
  \BibitemOpen
  \bibfield  {author} {\bibinfo {author} {\bibfnamefont {A.~S.}\ \bibnamefont
  {Alexandrov}}\ and\ \bibinfo {author} {\bibfnamefont {J.~T.}\ \bibnamefont
  {Devreese}},\ }\href@noop {} {\emph {\bibinfo {title} {Advances in polaron
  physics}}}\ (\bibinfo  {publisher} {Springer},\ \bibinfo {year}
  {2010})\BibitemShut {NoStop}%
\bibitem [{\citenamefont {Dykman}\ and\ \citenamefont
  {Rashba}(2015)}]{dykman2015roots}%
  \BibitemOpen
  \bibfield  {author} {\bibinfo {author} {\bibfnamefont {M.~I.}\ \bibnamefont
  {Dykman}}\ and\ \bibinfo {author} {\bibfnamefont {E.~I.}\ \bibnamefont
  {Rashba}},\ }\href@noop {} {\bibfield  {journal} {\bibinfo  {journal}
  {Physics Today}\ }\textbf {\bibinfo {volume} {68}},\ \bibinfo {pages} {10}
  (\bibinfo {year} {2015})}\BibitemShut {NoStop}%
\bibitem [{\citenamefont {Andrei}(1984)}]{PhysRevLett.52.1449}%
  \BibitemOpen
  \bibfield  {author} {\bibinfo {author} {\bibfnamefont {E.~Y.}\ \bibnamefont
  {Andrei}},\ }\href {\doibase 10.1103/PhysRevLett.52.1449} {\bibfield
  {journal} {\bibinfo  {journal} {Phys. Rev. Lett.}\ }\textbf {\bibinfo
  {volume} {52}},\ \bibinfo {pages} {1449} (\bibinfo {year}
  {1984})}\BibitemShut {NoStop}%
\bibitem [{\citenamefont {Johansson}\ and\ \citenamefont
  {Stafstr\"om}(2004)}]{PhysRevB.69.235205}%
  \BibitemOpen
  \bibfield  {author} {\bibinfo {author} {\bibfnamefont {A.~A.}\ \bibnamefont
  {Johansson}}\ and\ \bibinfo {author} {\bibfnamefont {S.}~\bibnamefont
  {Stafstr\"om}},\ }\href {\doibase 10.1103/PhysRevB.69.235205} {\bibfield
  {journal} {\bibinfo  {journal} {Phys. Rev. B}\ }\textbf {\bibinfo {volume}
  {69}},\ \bibinfo {pages} {235205} (\bibinfo {year} {2004})}\BibitemShut
  {NoStop}%
\bibitem [{\citenamefont {Ge}\ \emph {et~al.}(1998)\citenamefont {Ge},
  \citenamefont {Wong}, \citenamefont {Lingle}, \citenamefont {McNeill},
  \citenamefont {Gaffney},\ and\ \citenamefont {Harris}}]{ge1998femtosecond}%
  \BibitemOpen
  \bibfield  {author} {\bibinfo {author} {\bibfnamefont {N.-H.}\ \bibnamefont
  {Ge}}, \bibinfo {author} {\bibfnamefont {C.}~\bibnamefont {Wong}}, \bibinfo
  {author} {\bibfnamefont {R.}~\bibnamefont {Lingle}}, \bibinfo {author}
  {\bibfnamefont {J.}~\bibnamefont {McNeill}}, \bibinfo {author} {\bibfnamefont
  {K.}~\bibnamefont {Gaffney}}, \ and\ \bibinfo {author} {\bibfnamefont
  {C.}~\bibnamefont {Harris}},\ }\href@noop {} {\bibfield  {journal} {\bibinfo
  {journal} {Science}\ }\textbf {\bibinfo {volume} {279}},\ \bibinfo {pages}
  {202} (\bibinfo {year} {1998})}\BibitemShut {NoStop}%
\bibitem [{\citenamefont {Gaal}\ \emph {et~al.}(2007)\citenamefont {Gaal},
  \citenamefont {Kuehn}, \citenamefont {Reimann}, \citenamefont {Woerner},
  \citenamefont {Elsaesser},\ and\ \citenamefont {Hey}}]{gaal2007}%
  \BibitemOpen
  \bibfield  {author} {\bibinfo {author} {\bibfnamefont {P.}~\bibnamefont
  {Gaal}}, \bibinfo {author} {\bibfnamefont {W.}~\bibnamefont {Kuehn}},
  \bibinfo {author} {\bibfnamefont {K.}~\bibnamefont {Reimann}}, \bibinfo
  {author} {\bibfnamefont {M.}~\bibnamefont {Woerner}}, \bibinfo {author}
  {\bibfnamefont {T.}~\bibnamefont {Elsaesser}}, \ and\ \bibinfo {author}
  {\bibfnamefont {R.}~\bibnamefont {Hey}},\ }\href@noop {} {\bibfield
  {journal} {\bibinfo  {journal} {Nature}\ }\textbf {\bibinfo {volume} {450}},\
  \bibinfo {pages} {1210} (\bibinfo {year} {2007})}\BibitemShut {NoStop}%
\bibitem [{\citenamefont {Monarkha}\ and\ \citenamefont
  {Kono}(2005)}]{MKDimple}%
  \BibitemOpen
  \bibfield  {author} {\bibinfo {author} {\bibfnamefont {Y.~P.}\ \bibnamefont
  {Monarkha}}\ and\ \bibinfo {author} {\bibfnamefont {K.}~\bibnamefont
  {Kono}},\ }\href {\doibase 10.1143/JPSJ.74.960} {\bibfield  {journal}
  {\bibinfo  {journal} {J. Phys. Soc. Jpn.}\ }\textbf {\bibinfo {volume}
  {74}},\ \bibinfo {pages} {960} (\bibinfo {year} {2005})}\BibitemShut
  {NoStop}%
\bibitem [{\citenamefont {Dykman}\ and\ \citenamefont
  {Rubo}(1997)}]{DykmanRubo}%
  \BibitemOpen
  \bibfield  {author} {\bibinfo {author} {\bibfnamefont {M.~I.}\ \bibnamefont
  {Dykman}}\ and\ \bibinfo {author} {\bibfnamefont {Y.~G.}\ \bibnamefont
  {Rubo}},\ }\href {\doibase 10.1103/PhysRevLett.78.4813} {\bibfield  {journal}
  {\bibinfo  {journal} {Phys. Rev. Lett.}\ }\textbf {\bibinfo {volume} {78}},\
  \bibinfo {pages} {4813} (\bibinfo {year} {1997})}\BibitemShut {NoStop}%
\bibitem [{\citenamefont {Vinen}(1999)}]{Vinen}%
  \BibitemOpen
  \bibfield  {author} {\bibinfo {author} {\bibfnamefont {W.~F.}\ \bibnamefont
  {Vinen}},\ }\href@noop {} {\bibfield  {journal} {\bibinfo  {journal} {J.
  Phys.: Condens. Matter}\ }\textbf {\bibinfo {volume} {11}},\ \bibinfo {pages}
  {9709} (\bibinfo {year} {1999})}\BibitemShut {NoStop}%
\bibitem [{\citenamefont {Kristensen}\ \emph {et~al.}(1996)\citenamefont
  {Kristensen}, \citenamefont {Djerfi}, \citenamefont {Fozooni}, \citenamefont
  {Lea}, \citenamefont {Richardson}, \citenamefont {Santrich-Badal},
  \citenamefont {Blackburn},\ and\ \citenamefont {van~der
  Heijden}}]{Kristensen1996}%
  \BibitemOpen
  \bibfield  {author} {\bibinfo {author} {\bibfnamefont {A.}~\bibnamefont
  {Kristensen}}, \bibinfo {author} {\bibfnamefont {K.}~\bibnamefont {Djerfi}},
  \bibinfo {author} {\bibfnamefont {P.}~\bibnamefont {Fozooni}}, \bibinfo
  {author} {\bibfnamefont {M.~J.}\ \bibnamefont {Lea}}, \bibinfo {author}
  {\bibfnamefont {P.~J.}\ \bibnamefont {Richardson}}, \bibinfo {author}
  {\bibfnamefont {A.}~\bibnamefont {Santrich-Badal}}, \bibinfo {author}
  {\bibfnamefont {A.}~\bibnamefont {Blackburn}}, \ and\ \bibinfo {author}
  {\bibfnamefont {R.~W.}\ \bibnamefont {van~der Heijden}},\ }\href {\doibase
  10.1103/PhysRevLett.77.1350} {\bibfield  {journal} {\bibinfo  {journal}
  {Phys. Rev. Lett.}\ }\textbf {\bibinfo {volume} {77}},\ \bibinfo {pages}
  {1350} (\bibinfo {year} {1996})}\BibitemShut {NoStop}%
\bibitem [{\citenamefont {Shirahama}\ and\ \citenamefont
  {Kono}(1995)}]{PhysRevLett.74.781}%
  \BibitemOpen
  \bibfield  {author} {\bibinfo {author} {\bibfnamefont {K.}~\bibnamefont
  {Shirahama}}\ and\ \bibinfo {author} {\bibfnamefont {K.}~\bibnamefont
  {Kono}},\ }\href {\doibase 10.1103/PhysRevLett.74.781} {\bibfield  {journal}
  {\bibinfo  {journal} {Phys. Rev. Lett.}\ }\textbf {\bibinfo {volume} {74}},\
  \bibinfo {pages} {781} (\bibinfo {year} {1995})}\BibitemShut {NoStop}%
\bibitem [{\citenamefont {Shirahama}\ and\ \citenamefont
  {Kono}(1996)}]{Shirahama2}%
  \BibitemOpen
  \bibfield  {author} {\bibinfo {author} {\bibfnamefont {K.}~\bibnamefont
  {Shirahama}}\ and\ \bibinfo {author} {\bibfnamefont {K.}~\bibnamefont
  {Kono}},\ }\href {\doibase 10.1007/BF00754096} {\bibfield  {journal}
  {\bibinfo  {journal} {J. Low Temp. Phys.}\ }\textbf {\bibinfo {volume}
  {104}},\ \bibinfo {pages} {237} (\bibinfo {year} {1996})}\BibitemShut
  {NoStop}%
\bibitem [{\citenamefont {Ikegami}\ \emph {et~al.}(2009)\citenamefont
  {Ikegami}, \citenamefont {Akimoto},\ and\ \citenamefont
  {Kono}}]{Ikegami2009}%
  \BibitemOpen
  \bibfield  {author} {\bibinfo {author} {\bibfnamefont {H.}~\bibnamefont
  {Ikegami}}, \bibinfo {author} {\bibfnamefont {H.}~\bibnamefont {Akimoto}}, \
  and\ \bibinfo {author} {\bibfnamefont {K.}~\bibnamefont {Kono}},\ }\href
  {\doibase 10.1103/PhysRevLett.102.046807} {\bibfield  {journal} {\bibinfo
  {journal} {Phys. Rev. Lett.}\ }\textbf {\bibinfo {volume} {102}},\ \bibinfo
  {pages} {046807} (\bibinfo {year} {2009})}\BibitemShut {NoStop}%
\bibitem [{\citenamefont {Hecht}(2012)}]{hecht2012new}%
  \BibitemOpen
  \bibfield  {author} {\bibinfo {author} {\bibfnamefont {F.}~\bibnamefont
  {Hecht}},\ }\href@noop {} {\bibfield  {journal} {\bibinfo  {journal} {J.
  Numer. Math.}\ }\textbf {\bibinfo {volume} {20}},\ \bibinfo {pages} {251}
  (\bibinfo {year} {2012})}\BibitemShut {NoStop}%
\bibitem [{\citenamefont {Beysengulov}\ \emph {et~al.}(2016)\citenamefont
  {Beysengulov}, \citenamefont {Rees}, \citenamefont {Lysogorskiy},
  \citenamefont {Galiullin}, \citenamefont {Vazjukov}, \citenamefont
  {Tayurskii},\ and\ \citenamefont {Kono}}]{beysengulov2016structural}%
  \BibitemOpen
  \bibfield  {author} {\bibinfo {author} {\bibfnamefont {N.~R.}\ \bibnamefont
  {Beysengulov}}, \bibinfo {author} {\bibfnamefont {D.~G.}\ \bibnamefont
  {Rees}}, \bibinfo {author} {\bibfnamefont {Y.}~\bibnamefont {Lysogorskiy}},
  \bibinfo {author} {\bibfnamefont {N.~K.}\ \bibnamefont {Galiullin}}, \bibinfo
  {author} {\bibfnamefont {A.~S.}\ \bibnamefont {Vazjukov}}, \bibinfo {author}
  {\bibfnamefont {D.~A.}\ \bibnamefont {Tayurskii}}, \ and\ \bibinfo {author}
  {\bibfnamefont {K.}~\bibnamefont {Kono}},\ }\href@noop {} {\bibfield
  {journal} {\bibinfo  {journal} {J. Low Temp. Phys.}\ }\textbf {\bibinfo
  {volume} {182}},\ \bibinfo {pages} {28} (\bibinfo {year} {2016})}\BibitemShut
  {NoStop}%
\bibitem [{Sup()}]{Supp}%
  \BibitemOpen
  \href@noop {} {\ }\bibinfo {note} {See Supplemental Material at for details
  of the experimental method, electrostatic modelling and ac transport
  measurements used to characterise the device.}\BibitemShut {Stop}%
\bibitem [{\citenamefont {Cs\'athy}\ \emph {et~al.}(2007)\citenamefont
  {Cs\'athy}, \citenamefont {Tsui}, \citenamefont {Pfeiffer},\ and\
  \citenamefont {West}}]{Csathy}%
  \BibitemOpen
  \bibfield  {author} {\bibinfo {author} {\bibfnamefont {G.~A.}\ \bibnamefont
  {Cs\'athy}}, \bibinfo {author} {\bibfnamefont {D.~C.}\ \bibnamefont {Tsui}},
  \bibinfo {author} {\bibfnamefont {L.~N.}\ \bibnamefont {Pfeiffer}}, \ and\
  \bibinfo {author} {\bibfnamefont {K.~W.}\ \bibnamefont {West}},\ }\href
  {\doibase 10.1103/PhysRevLett.98.066805} {\bibfield  {journal} {\bibinfo
  {journal} {Phys. Rev. Lett.}\ }\textbf {\bibinfo {volume} {98}},\ \bibinfo
  {pages} {066805} (\bibinfo {year} {2007})}\BibitemShut {NoStop}%
\bibitem [{\citenamefont {Persson}(2000)}]{persson2000sliding}%
  \BibitemOpen
  \bibfield  {author} {\bibinfo {author} {\bibfnamefont {B.~N.~J.}\
  \bibnamefont {Persson}},\ }\href@noop {} {\emph {\bibinfo {title} {Sliding
  friction: physical principles and applications}}},\ Vol.~\bibinfo {volume}
  {1}\ (\bibinfo  {publisher} {Springer Science \& Business Media},\ \bibinfo
  {year} {2000})\BibitemShut {NoStop}%
\bibitem [{\citenamefont {Ikegami}\ \emph {et~al.}(2012)\citenamefont
  {Ikegami}, \citenamefont {Akimoto}, \citenamefont {Rees},\ and\ \citenamefont
  {Kono}}]{PhysRevLett.109.236802}%
  \BibitemOpen
  \bibfield  {author} {\bibinfo {author} {\bibfnamefont {H.}~\bibnamefont
  {Ikegami}}, \bibinfo {author} {\bibfnamefont {H.}~\bibnamefont {Akimoto}},
  \bibinfo {author} {\bibfnamefont {D.~G.}\ \bibnamefont {Rees}}, \ and\
  \bibinfo {author} {\bibfnamefont {K.}~\bibnamefont {Kono}},\ }\href {\doibase
  10.1103/PhysRevLett.109.236802} {\bibfield  {journal} {\bibinfo  {journal}
  {Phys. Rev. Lett.}\ }\textbf {\bibinfo {volume} {109}},\ \bibinfo {pages}
  {236802} (\bibinfo {year} {2012})}\BibitemShut {NoStop}%
\bibitem [{\citenamefont {G.~Rees}\ \emph {et~al.}(2013)\citenamefont
  {G.~Rees}, \citenamefont {Ikegami},\ and\ \citenamefont
  {Kono}}]{rees2013reentrant}%
  \BibitemOpen
  \bibfield  {author} {\bibinfo {author} {\bibfnamefont {D.}~\bibnamefont
  {G.~Rees}}, \bibinfo {author} {\bibfnamefont {H.}~\bibnamefont {Ikegami}}, \
  and\ \bibinfo {author} {\bibfnamefont {K.}~\bibnamefont {Kono}},\ }\href@noop
  {} {\bibfield  {journal} {\bibinfo  {journal} {J. Phys. Soc. Jpn.}\ }\textbf
  {\bibinfo {volume} {82}},\ \bibinfo {pages} {124602} (\bibinfo {year}
  {2013})}\BibitemShut {NoStop}%
\bibitem [{\citenamefont {Vanossi}\ \emph {et~al.}(2013)\citenamefont
  {Vanossi}, \citenamefont {Manini}, \citenamefont {Urbakh}, \citenamefont
  {Zapperi},\ and\ \citenamefont {Tosatti}}]{RevModPhys.85.529}%
  \BibitemOpen
  \bibfield  {author} {\bibinfo {author} {\bibfnamefont {A.}~\bibnamefont
  {Vanossi}}, \bibinfo {author} {\bibfnamefont {N.}~\bibnamefont {Manini}},
  \bibinfo {author} {\bibfnamefont {M.}~\bibnamefont {Urbakh}}, \bibinfo
  {author} {\bibfnamefont {S.}~\bibnamefont {Zapperi}}, \ and\ \bibinfo
  {author} {\bibfnamefont {E.}~\bibnamefont {Tosatti}},\ }\href {\doibase
  10.1103/RevModPhys.85.529} {\bibfield  {journal} {\bibinfo  {journal} {Rev.
  Mod. Phys.}\ }\textbf {\bibinfo {volume} {85}},\ \bibinfo {pages} {529}
  (\bibinfo {year} {2013})}\BibitemShut {NoStop}%
\bibitem [{\citenamefont {Lee}\ \emph {et~al.}(2010)\citenamefont {Lee},
  \citenamefont {Li}, \citenamefont {Kalb}, \citenamefont {Liu}, \citenamefont
  {Berger}, \citenamefont {Carpick},\ and\ \citenamefont
  {Hone}}]{lee2010frictional}%
  \BibitemOpen
  \bibfield  {author} {\bibinfo {author} {\bibfnamefont {C.}~\bibnamefont
  {Lee}}, \bibinfo {author} {\bibfnamefont {Q.}~\bibnamefont {Li}}, \bibinfo
  {author} {\bibfnamefont {W.}~\bibnamefont {Kalb}}, \bibinfo {author}
  {\bibfnamefont {X.-Z.}\ \bibnamefont {Liu}}, \bibinfo {author} {\bibfnamefont
  {H.}~\bibnamefont {Berger}}, \bibinfo {author} {\bibfnamefont {R.~W.}\
  \bibnamefont {Carpick}}, \ and\ \bibinfo {author} {\bibfnamefont
  {J.}~\bibnamefont {Hone}},\ }\href@noop {} {\bibfield  {journal} {\bibinfo
  {journal} {Science}\ }\textbf {\bibinfo {volume} {328}},\ \bibinfo {pages}
  {76} (\bibinfo {year} {2010})}\BibitemShut {NoStop}%
\bibitem [{\citenamefont {Lee}\ \emph {et~al.}(2015)\citenamefont {Lee},
  \citenamefont {Kim}, \citenamefont {Kim},\ and\ \citenamefont
  {Jhe}}]{lee2015noncontact}%
  \BibitemOpen
  \bibfield  {author} {\bibinfo {author} {\bibfnamefont {M.}~\bibnamefont
  {Lee}}, \bibinfo {author} {\bibfnamefont {B.}~\bibnamefont {Kim}}, \bibinfo
  {author} {\bibfnamefont {J.}~\bibnamefont {Kim}}, \ and\ \bibinfo {author}
  {\bibfnamefont {W.}~\bibnamefont {Jhe}},\ }\href@noop {} {\bibfield
  {journal} {\bibinfo  {journal} {Nature communications}\ }\textbf {\bibinfo
  {volume} {6}} (\bibinfo {year} {2015})}\BibitemShut {NoStop}%
\bibitem [{\citenamefont {Bylinskii}\ \emph {et~al.}(2015)\citenamefont
  {Bylinskii}, \citenamefont {Gangloff},\ and\ \citenamefont
  {Vuleti\'c}}]{Bylinskii}%
  \BibitemOpen
  \bibfield  {author} {\bibinfo {author} {\bibfnamefont {A.}~\bibnamefont
  {Bylinskii}}, \bibinfo {author} {\bibfnamefont {D.}~\bibnamefont {Gangloff}},
  \ and\ \bibinfo {author} {\bibfnamefont {V.}~\bibnamefont {Vuleti\'c}},\
  }\href@noop {} {\bibfield  {journal} {\bibinfo  {journal} {Science}\ }\textbf
  {\bibinfo {volume} {348}},\ \bibinfo {pages} {1115} (\bibinfo {year}
  {2015})}\BibitemShut {NoStop}%
\end{thebibliography}%

\bibliographystyle{apsrev4-1}

\clearpage

\section{Supplemental - Materials and Methods}

The sample was fabricated on a silicon wafer using optical lithography. Two gold layers, each 80 nm thick, were separated by an insulating layer of hard-baked photoresist. The metal layers were patterned to form the different electrodes. The widths of the gaps between the electrodes were approximately 0.5 $\mu$m in the lower layer and 3 $\mu$m in the upper layer. After lift-off of the second layer, the photoresist that was not covered by metal was removed by etching to form the channel structures. The reservoirs comprised 25 microchannels, each 20 $\mu$m wide and 700 $\mu$m long, connected in parallel.

The samples were wire bonded to a ceramic chip carrier that was then mounted in a hermetically sealed copper sample cell attached to the mixing chamber of a dilution refrigerator. $^4$He was condensed into the sample cell until the surface of the superfluid was approximately 0.5 mm below the sample, allowing the microchannels to fill by the capillary action of the superfluid film. A 100 ms voltage pulse was applied to a tungsten filament placed several mm above the sample to induce the thermal emission of electrons. The charging procedure was performed at 0.9 K, at which temperature the electrons thermalise rapidly due to collisions with He gas atoms before being trapped on the surface. 

Ac transport measurements, in which the driving force applied to the SSE can be made very small, were performed by superimposing a small ac voltage $V_{in}$ on the Left Reservoir electrode and measuring the ac current $I_{ac}$ induced in the Right Reservoir electrode, using the current preamplifier and a lock-in amplifier. The ac frequency was 89.2 kHz.

\section{Supplemental - Ac transport measurements and Finite Element Modelling}

Before investigating the dc transport, we performed ac transport measurements, and used the results to build an electrostatic model of the device with the aid of finite element modelling (FEM). We first measured the Split-Gate voltage threshold of current flow $V_{sg}^{th}$ for different values of $V_{bg}$. The results of the measurement are shown in Fig. 1(a) for $V_{gu}=-0.2$ V, $V_{lr}=V_{rr}=0$ V and $V_{in}=3$ mV$_{pp}$. For such small driving voltage, the system remains close to equilibrium during the measurement. For increasingly positive $V_{bg}$, a more negative $V_{sg}$ is required to `pinch-off' the current. At this threshold, where electrons enter the CM forming a single row along the centre of the channel, two energetic conditions must be satisfied. Firstly, the potential at the centre of the CM, $V_c$, must be more positive than the electrostatic potential of the electron system $V_e$. Were the electron system to behave as a charge continuum, the satisfaction of this condition alone would result in current flow. However, because the electron system is granular in nature, a second condition arises, that the charging energy required to populate the CM with electrons must be compensated for. We consider that this condition is approximately satisfied when the number of electrons allowed in a square section of the effective CM area $S = w_e^2$ is equal to 1 or, equivalently, that the separation between the electrons along the centre of the CM is comparable to $w_e$.

We then used a finite element modelling software package[20] to solve Poisson's equation for the electrostatic potential $\phi$ at all points in the CM cross section, according to the geometry and bias conditions of the sample. Without electrons, the potential at the centre of the CM was given by $V_{c}=\alpha V_{bg}+\beta V_{sg}$, where $\alpha$ and $\beta$ are coupling constants. (The influence of the other electrodes is negligible as they are strongly screened.) The electron system was then modelled as a continuous charge sheet of width $w_e$ and potential $V_e$. The width was determined, for varying electrode bias conditions, by iteratively calculating the value of $w_e$ for which the electric field at the edge of the sheet, in the plane of the helium surface, was zero.  The surface charge density distribution across the electron sheet $n_s(y)$ was then found by measuring the difference between the electric field above and below the electron sheet, using the expression

\begin{eqnarray}  \label{eq:ChargeDensity}
\frac{\partial \phi}{\partial z}\biggr\vert_{above} - \varepsilon_{He}\frac{\partial \phi}{\partial z}\biggr\vert_{below}  = \frac{-en_s(y)}{\varepsilon_0}~~,
\end{eqnarray}

where $\varepsilon_0$ is the permittivity of free space and $\varepsilon_{He}=1.056$ is the dielectric constant of liquid He. The average areal density $n_s$ was given by the average value of $n_s(y)$ whilst the linear electron density $n_l$ in the CM was evaluated numerically using the expression

\begin{eqnarray}  \label{eq:LinearDensity}
n_l = \int_{\frac{-w_e}{2}}^{\frac{w_e}{2}} n_s(y) dy~~.
\end{eqnarray}

\noindent The number of electron rows in the CM, $N_y$, was then estimated as $N_y=\sqrt{w_en_l}$. Given our fabrication procedure, we expect a channel depth of approximately 2 $\mu$m. Using a channel depth of 2.2 $\mu$m, for which $\alpha=0.6$ and $\beta=0.4$, and the value $V_e=-0.13$ V we found an excellent agreement between the voltage values for which our model predicts $N_y=1$ and the current threshold line, as shown in Fig. 1(a).    

\begin{figure} 
\begin{centering}
  \includegraphics[angle=0,width=0.45\textwidth]{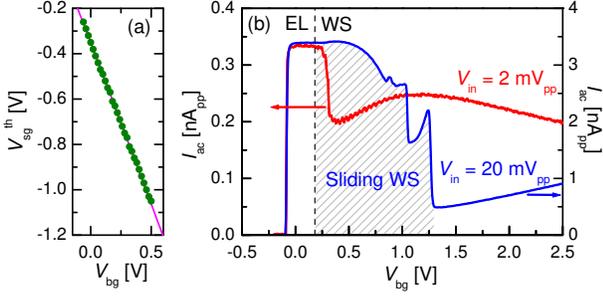}
  \caption{Results of ac transport measurements. (a) $V_{sg}^{th}$ against $V_{bg}$ for $V_{lr}=V_{rr}=0$ V, $V_{gu}=-0.2$ V, $V_{in}=3$ mV$_{pp}$ and $T=0.6$ K. The solid line shows the result of FEM modelling for $N_y=1$. (b) $I_{ac}$ against $V_{bg}$ for $V_{gu}=V_{sg}=-0.2$ V and $V_{in}=2$ and 20 mV$_{pp}$. The axes for the two measurements are scaled by a factor of 10 to allow clear comparison. The transition between the electron liquid (EL) and WS phases occurs when $V_{bg}\approx 0.18$ V. For $V_{in}=20$ mV$_{pp}$ the WS `slides' along the Helium surface before becoming pinned at $V_{bg}=1.30$ V.\label{Fig:1}}
  \end{centering}
\end{figure}

In Fig. 1(b) we show the dependence of $I_{ac}$ on $V_{bg}$ for two values of $V_{in}$. Here $T=0.6$ K and the melting of the classical WS occurs for $n_{s}\approx7.1\times10^8$ cm$^{-2}$. Our FEM model gives this value for $n_{s}$ when $V_{bg}=0.18$ V. For $V_{in}=2$ mV$_{pp}$, the current drops abruptly close to this value as the WS couples with the DL. (In this regime, we also note the appearance of current oscillations as the number of electron rows in the CM, and so the commensurability of the electron lattice with the confinement, is modulated. These effects will be described in detail elsewhere). For $V_{in}=20$ mV$_{pp}$, the current remains high as the WS slides along the Helium surface, before the pinning with the DL occurs at $V_{bg}=1.30$ V. In the sliding WS regime a current peak is observed immediately before the SSE are pinned by the DL. For $V_{in}=20$ mV$_{pp}$, the root-mean-square rate of change of voltage is 3.9 mV/$\mu$s, comparable with the 2.5 mV/$\mu$s used in Fig. 3 in the main text. We therefore conclude that the current peak observed in the ac transport measurement arises due to WS sliding during each ac voltage cycle, resulting in a large signal component recorded at the ac reference frequency. Other resonances weakly observed in the ac measurement at lower values of $V_{bg}$ occur when multiple sliding events occur during each cycle.

\end{document}